\documentclass[11pt]{article}
\usepackage{amsmath}
\usepackage{amsfonts}
\usepackage{amssymb}
\usepackage{graphicx}
\usepackage{subfigure}
\usepackage{graphicx}
\usepackage{dcolumn}
\usepackage{bm}
\usepackage[all]{xy}
\usepackage{setspace}
\graphicspath{{images/}}
\usepackage{color}             
\definecolor{Dblue}{rgb}{0.0,0.0,0.4}
\definecolor{Dgreen}{rgb}{0.0,0.5,0.0}
\definecolor{Dred}{rgb}{0.7,0.0,0.0}
\definecolor{Lyellow}{rgb}{1.0,1.0,0.8}
\definecolor{white}{rgb}{1.0,1.0,1.0}
\usepackage[table]{xcolor}

cm \headheight=0. mm

\def\be{\begin{equation}}
 \def\ee{\end{equation}}
\def\bea{\begin{eqnarray}}
\def\eea{\end{eqnarray}}

\begin{document}
\begin{center}
\LARGE {Non-local scalar fields inflationary mechanism in light of Planck $2013$}
\end{center}
\begin{center}
{\bf $^{a} $Haidar Sheikhahmadi\footnote{h.sh.ahmadi@gmail.com}\\
{\bf $^{a} $Soheyla Ghorbani\footnote{ghorbani.soheyla@ymail.com}\\
{\bf $^{a},^{b}$Khaled Saaidi\footnote{khaledsaeidi@gmail.com}, \footnote{kh.saeidi@stpok.ir}\\
{\it $^a$Department of Physics, Faculty of Science, University of Kurdistan,  Sanandaj, Iran.\\}
{\it $^b$ Science and Technology Park of Kurdistan, Sanandaj, Iran.}}}}\\
\end{center}
 \vskip 1cm
\begin{abstract}
{A generalization of the canonical and non-canonical theory of inflation is introduced in which the kinetic energy term in action is written as non-local term. The inflationary universe within the framework of  considering this non-locality will be studied. To investigate the effects of non-locality on the inflationary parameters we consider two well known models of inflationary scenario includes of chaotic and exponential inflation proposals. For such scenarios  some important parameters include slow roll parameters, scalar and tensor power spectra, spectral indices, the tensor-to-scalar ratio and so on for both mentioned models, chaotic and exponential inflationary scenarios, will be calculated. Also the Hamilton-Jacobi formalism, as an easiest way to study the effect of perturbation based on e-folding number $N$, to investigate inflationary attractors will be used. The free theoretical parameters of this model will be compared with observations by means of  Planck $2013$, $WMAP9+eCMB+BAO+H_0$ data sets in addition to $BICEP2$ data surveying. It will be shown that our theoretical results are in acceptable range in comparison to observations. For instance the tensor-to-scalar ratio for exponential potential, by considering $BICEP2$ is in best agreement in comparison with chaotic inflation.}
\end{abstract}

\newpage
\section{Introduction}
Recent observational data, and  related mechanisms, includes of Cosmic Microwave Background  (CMB) \cite{CMB_th_1, CMB2}, Supernovae type Ia (SNeIa) \cite{ch1:7, ch1:8}, Baryonic Acoustic Oscillations (BAO) \cite{BAO_th_Eisenstein, BAO_th_Shoji}, Observational Hubble Data (OHD) \cite{FarooqRatra, H0_1}, Sloan Digital Sky Survey (SDSS) \cite{ch1:10, SD2}, and Wilkinson Microwave Anisotropy Probe (WMAP)~\cite{ch1:9,WM2}, show that the Universe is undergoing an accelerated expansion epoch.  To justify this ambiguity, scientists propose that the Universe should dominated with an ambitious form of matter namely dark energy. The above mentioned observations also suggest that the components of the universe are dark enery $73\%$, dark matter $23\%$ and baryonic matter only $4\%$. For present epoch of the Universe evolution radiation can be ignored. The interesting and surprising fact is that the nature and origin of dark energy are unknown for people up to now. The evidences indicate that the best candidate for dark energy is cosmological constant $\Lambda$, \cite{3, 3a, 3d,  3b, 3e}. It is well known this important candidate of dark energy suffers two well known problems namely fine tuning problem and coincidence problem, which the first one refers to the difference of the theoretical anticipation of the energy density of $\Lambda$ in comparison to observations and the latter asks, why the ratio of energy densities of dark energy and dark matter in present epoch is so closed to unity?\cite{4}. These two problems convince scientists to make other proposals
to justify the existence and behaviour of dark energy. Amongst such proposals people interestingly investigate scalar field scenarios. This attractive and powerful proposal includes of a wide range of different models as Brans-Dicke {\cite{brans1, brans2}}, quintessence {\cite{5, 5a, 5c, 5b}}, k-essence {\cite{6, 6a, 6b}}, tachyon {\cite{7}}, phantom {\cite{8, 8a, 8b}}, quintom {\cite{9, 9a, 9b}}, and chameleon{\cite{11, 11a, 10, ch18, ch19, ch20, ch1:20a, ch1:18}}.\\
Recently investigation the inflation using scalar fields attract more attention, for instance canonical and local inflation \cite{ca1, ca2, ca3}, non canonical and local inflation \cite{nca1, nca2, nca3}, brain inflation with canonical and non canonical inflation {\cite{bca1, bca2}} and so on. It should be stressed \cite{refine} where is devoted to investigate the effects of a non-canonical scalar field on the inflation evolution,  motivated us to study the effects of a non local scalar field on the inflation scenario. Let us refer to the origin of a non-local mechanism. For instance it is obvious that the modified gravity models proposed to resolve some open problems which standard model was faced to them. Amongst different proposals to get rid of ultraviolet behavior and also the renormalizability problem one maybe refer to the addition of higher derivative terms of the Ricci scalar to the action, but it is obvious that such mechanism produces the ghosts{\cite{Impo}}. To overcome such ghosts in this model, people usually consider non-local mechanism{\cite{Impo01, Impo03}}. Also it should be noted the main source of non-local mechanism comes from string theory{\cite{Impo02}}. It is notable generalization of Yang-Mills theories instead of using an addition auxiliary field to the gauge field, one can use the non-local action and cause to useful results in gauge field theories scenario{\cite{Xo}}. We should emphasise for such mechanism in comparison to the canonical and local scalar field approach, by considering observations includes of   Planck $2013$ {\cite{PL01}}, $WMAP9+eCMB+BAO+H_0$ data sets in addition to $BICEP2$ data surveying{\cite{Ade01, Ade02}}, our results are in good agreement.\\
The paper has been planned in the following form: In Sec.1\, which includes of above discussions we have introduction. In Sec.2\, the general formalism and related equations are investigated. The slow roll parameters for some steep potential like Ratra-Peebles potential is obtained. Using The Hamilton-Jacobi equation the inflationary  attractors are discussed, also time evolution of the first slow roll parameter is obtained. In this section also scalar and tensor power spectra for steep potentials will be achieved. In Sec.3\, using power law and exponential potential the inflationary behavior of the Universe will be investigated. As important free parameters, scalar and tensor spectral indices and also tensor-to-scalar ratio will be attained. Also for some different e-foldings number the important free parameters of the model with observed values will be compared. At last Sec.4\, is devoted to conclusion and discussions.
\section{Non-local formalism and General Framework}
To investigate the effects of each component of space time on other points, one can consider a non-local action as
\begin{equation}\label{1-action}
S = F[{g_{\mu \upsilon }},\,\phi ] + \int {{d^4}x\sqrt { - g}\big (\frac{R}{{2\kappa }}}  + {L_m} - V(\phi )\big),
\end{equation}
where
$F[{g_{\mu \upsilon }},\,\phi ] = F[\int {{d^4}x\sqrt { - g} (}  - \frac{1}{2}{g^{_{\mu \upsilon }}}{\partial _\mu }\phi {\partial _\upsilon }\phi )],$ $R$ is Ricci scalar, ${L_m}$ is the Lagrangian of matter and $V(\phi )$ is an arbitrary potential. Variation action (\ref{1-action}) with respect to (w.r.t) the metric tensor gives
\begin{equation}\label{2-Ei-tensor}
{G_{\mu \upsilon }} = \kappa (T_{\mu \upsilon }^m + T_{\mu \upsilon }^\phi ),
\end{equation}
where $\kappa=8\pi G$, $T_{\mu \upsilon }^m = \frac{{ - 2}}{{\sqrt { - g} }}\frac{{\delta (\sqrt { - g} {L_m})}}{{\delta {g^{\mu \upsilon }}}}$ and
\begin{equation}\label{3-T_phi-tensor}
T_{\mu \upsilon }^\phi  = \tilde F(\phi )({\partial _\mu }\phi {\partial _\upsilon }\phi  - \frac{1}{2}{\partial _\eta }\phi {\partial ^\eta }\phi ) - {g_{\mu \upsilon }}V(\phi ),
\end{equation}
here $\tilde F(\phi _{0})=\delta F/ \delta K$, $\phi _{0}$ can satisfy the evolution equation of the scalar field and $$K=\int {{d^4}x\sqrt { - g} (}  - \frac{1}{2}{g^{_{\mu \upsilon }}}{\partial _\mu }\phi {\partial _\upsilon }\phi ).$$ Also from non-locality it is well known that, although $\tilde F(\phi _{0})$ is a constant, but it gets different values for different $\phi_{0}$, namely $\tilde F(\phi _{0})\neq\tilde F(\phi _{0}^{\prime}).$
Calculating $00$ component of Eq.(\ref{2-Ei-tensor}) one obtains the Friedmann equation as
\begin{equation}\label{4-G00}
{H^2} = \kappa ({\rho _m} + {\rho _\phi }),
\end{equation}
where ${\rho _\phi } =  V(\phi )+{{\tilde F(\phi _{0}){{\dot \phi }^2}}}/{2}.$
One should note that during inflation energy density of matter can be omitted and it is energy density of scalar field derives the inflation evolution.
Considering $ii$ component of Eq.(\ref{2-Ei-tensor}) and also ${{\ddot a}}/{a} = { - 4\pi G}({P_\phi } + {\rho _\phi }/3)$ one finds
\begin{equation}\label{5-a-00}
\frac{{\ddot a}}{a} = \frac{{ - 8\pi G}}{3} \left[\tilde F(\phi_{0} ){{\dot \phi }^2} - V(\phi )\right],
\end{equation}
therefore $P_{\phi}$ can be obtained as
\begin{equation}\label{6-P}
{P_\phi } = \frac{{\tilde F(\phi_{0} ){{\dot \phi }^2}}}{2} - V(\phi ).
\end{equation}
Variation the action (\ref{1-action}) w.r.t scalar field gives the equation of motion of scalar field as
\begin{equation}\label{7-phi}
\ddot \phi  + 3H{{\dot \phi }} + \frac{{V'(\phi )}}{{\tilde F(\phi _{0})}} = 0,
\end{equation}
where over prime denotes $dV(\phi)/d\phi$.
\subsection{Slow roll parameters and inflation}
To investigate inflation scenario usually two important parameters namely slow roll parameters have crucial role. The first and second slow roll parameters are defined respectively as
\begin{equation}\label{8-1th slow roll}
\varepsilon  =  - \frac{{\dot H}}{{{H^2}}}
\end{equation}
and
\begin{equation}\label{8-2nd slow roll}
\delta  = \varepsilon  - \frac{{\dot \varepsilon }}{{2H\varepsilon }}.
\end{equation}
It is obviously seen that, from definition of deceleration parameter $q=-1-{{\dot H}}/{{{H^2}}}$, to justify inflationary period of the Universe the first slow roll parameter should smaller than unity $\varepsilon<1.$
Considering Eqs.(\ref{5-a-00}) and (\ref{8-1th slow roll}) for time evolution of Hubble parameter we have
\begin{equation}\label{9-Hubble evolution}
\dot H= - 4\pi G\tilde F(\phi_{0}){{\dot \phi }^2}.
\end{equation}
By means of Eq.(\ref{9-Hubble evolution}) and  Friedmnn equation one has
\begin{equation}\label{10-Friedmann02}
{H^2}(1 - \frac{\varepsilon }{3}) = \frac{{8\pi G}}{3}V(\phi ).
\end{equation}
For the second slow roll parameter by considering Eq.(\ref{8-1th slow roll}) one gets
\begin{equation}\label{11-2nd slow roll02}
\delta  =\frac{{ - \ddot H }}{{2H\dot H}}= \frac{{ - \ddot \phi }}{{H\dot \phi }},
\end{equation}
substituting above equation in Eq.(\ref{7-phi}), gives
\begin{equation}\label{12-2nd slow roll03}
3H\dot \phi (1 - \frac{\delta }{3}) =  - \frac{{V'(\phi )}}{{\tilde F(\phi_{0})}}.
\end{equation}
Therefore combination Eqs.(\ref{9-Hubble evolution}) and (\ref{10-Friedmann02}) and using the slow roll condition $\varepsilon<<1$ {\cite{refine} one has
\begin{equation}\label{13-1st slow rool 03}
\varepsilon  = \frac{3}{2}\,\frac{{\tilde F(\phi_{0}){{\dot \phi }^2}}}{{V(\phi )}}.
\end{equation}
Now by getting $\dot{\phi}$ from Eq.(\ref{12-2nd slow roll03}) and introduce it to relation (\ref{13-1st slow rool 03}) we attain
\begin{equation}\label{14-1st slow rool 04}
\varepsilon \simeq{\varepsilon_V} = {\Big[\frac{{V'(\phi ){M_{Pl}}}}{{V(\phi )\sqrt {2\tilde F(\phi_{0})} }}\Big]^2},
\end{equation}
in this formalism if $\tilde F(\phi_{0})=1$, the above quantity  reduces to the standard local and canonical expression of potential slow roll parameter
\begin{equation}\label{15-standard}
\varepsilon^{c, l}_V  = {\Big[\frac{{V'(\phi ){M_{Pl}}}}{{V(\phi )\sqrt {2} }}\Big]^2},
\end{equation}
where upper indices $c$ and $l$ refer to canonical and local respectively.

\subsection{Potentials and their role in inflation}
Whereas the relation between slow roll parameters and inflation potential is so important, we want to study the effect of some well known potentials to derive inflation. Amongst inflation potentials power law and exponential potentials attract more attentions. Therefore as the first example we consider Ratra-Peebles potential as
\begin{equation}\label{16-Ratra-Peebles}
V(\phi ) = {M^4}{M_{Pl}}^n{\phi ^{ - n}},
\end{equation}
where $M$ is a constant. Differentiate  the above equation w.r.t $\phi$ and using Eqs.(\ref{14-1st slow rool 04}) and (\ref{16-Ratra-Peebles}) we arrive
\begin{equation}\label{17-slow roll ratra- peebles}
\varepsilon = \frac{{{n^2}{M_{Pl}}^2}}{{2{\phi ^2}\tilde F(\phi_{0})}},
\end{equation}
to justify the condition that warm inflation can be occurred, scalar field should so small in comparison to Planck mass $\phi<<M_{Pl}$  and simultaneous $\phi \tilde F(\phi )>>M_{Pl}$. For our second example we consider exponential potential as
\begin{equation}\label{18-slow roll exponential}
V(\phi ) = {V_0}\Big({e^{\frac{{{M_{Pl}}}}{\phi }}} - 1\Big),
\end{equation}
where ${V_0}$ is a constant.  Calculating differentiation of Eq.(\ref{18-slow roll exponential}) and introduce it to Eq.(\ref{14-1st slow rool 04}) one has
\begin{equation}\label{19-slow roll exponential02}
{\varepsilon} = \frac{{{M_{Pl}}^4}}{{2{\phi ^4}\tilde F(\phi )}}{\Big(\frac{1}{{1 - {e^{ - \frac{{{M_{Pl}}}}{\phi }}}}}\Big)^2},
\end{equation}
for more discussion we consider some  cases which indicate the importance of a non-local model, if $\tilde F(\phi _{0})=1$ and $\phi  <  < {M_{Pl}}$ so $\varepsilon>>1$ and this model can not justify a warm inflation scenario, but as a second example if $\tilde F(\phi_{0} )\neq1$ and $\phi  <  < {M_{Pl}}$ and simultaneous $\phi \tilde F(\phi )>>M_{Pl}$ we have $\varepsilon<1$ and warm inflation can be happened. Therefore for both Ratra-Peebles and exponential potentials in some certain cases warm inflation can be justified in a non-local framework.

\subsection{Hamilto-Jacobi formalism and attractors}
It is well known to investigate the inflationary attractors, one easiest formalism is Hamilton-Jacobi mechanism. This formalism rises from Friedmann equation based on evolution of Hubble parameter $H(\phi)$. From Eq.(\ref{9-Hubble evolution}) we obtain
\begin{equation}\label{20-H Prime}
{H'}\equiv\frac{{dH}}{{d\phi }} = \frac{{{\rm{\dot H}}}}{{\dot \phi }} =  - 4\pi G\tilde F\dot \phi,
\end{equation}
using above equation and Eq.(\ref{4-G00}) one has
\begin{equation}\label{21-Hamilton Jacobi}
{{H'}^2} - \frac{3}{2}\frac{{\tilde F(\phi_{0})}}{{{M_{Pl}}^2}}{H^2} =  - \frac{1}{2}\frac{{\tilde F(\phi_{0})}}{{{M_{Pl}}^4}}V(\phi ),
\end{equation}
this equation is Hamilton-Jacobi equation for non-local frame work, it is reduce to local and canonical form if $\tilde F(\phi )=1$. To investigate two nearby trajectories in phase space one can use
\begin{equation}\label{22-H and delta H}
2H'\delta H' - \frac{3}{2}\frac{{\tilde F(\phi_{0})}}{{{M_{Pl}}^2}}2H\delta H = 0,
\end{equation}
integration above equation from $\phi_i$ to $\phi$ gives
\begin{equation}\label{23-delta H}
\delta H = \delta H({\phi _i})\exp \Big(\frac{3}{2}\frac{{\tilde F(\phi_{0})}}{{{M_{Pl}}^2}}\int\limits_{{\phi _i}}^{{\phi }} {d\phi (\frac{H}{{H'}})} \Big),
\end{equation}
where $\delta H({\phi _i})$ is related to perturbed Hubble parameter and $\phi_i$ is initial value for $\phi$. Using the definition of e-folding number
\begin{equation}\label{24-efolding}
(N - {N_i}) =  - \int\limits_{{\phi _i}}^\phi  {d\phi (\frac{H}{{\dot \phi }})},
\end{equation}
and Eq.(\ref{20-H Prime}), one has
\begin{equation}\label{25-efolding02}
N - {N_i} = 4\pi G \tilde F(\phi_{0})  \int \limits_{{\phi _i}}^\phi  \frac{H}{{{H'}}}d\phi.
\end{equation}
Substituting above equation in Eq.(\ref{23-delta H}) we find that
\begin{equation}\label{26-efolding03}
\delta H = \delta H({\phi _i})\exp (3(N - {N_i})),
\end{equation}
which is a rapid approach to inflationary attractor solution.
\subsection{Time evolution of the first slow roll parameter}
In this stage we can use the time evolution of the first slow roll parameter to attain the suitable potential to derive inflation. Combination of Eqs.(\ref{4-G00}), (\ref{8-1th slow roll}) and (\ref{9-Hubble evolution}) one gets
\begin{equation}\label{27-1th slow roll001}
\varepsilon  = 3\frac{{\tilde F(\phi ){{\dot \phi }^2}}}{{\tilde F(\phi_{0}){{\dot \phi }^2} + 2V(\phi )}},
\end{equation}
using definition of $\rho_{\phi}$ and above equation, we obtain the related inflation  potential as
\begin{equation}\label{28-potential}
V(\phi ) = {\rho _\phi }(1 - \frac{\varepsilon }{3}).
\end{equation}
Differentiation this equation w.r.t scalar field gives
\begin{equation}\label{29-potential prime}
V'(\phi ) = \frac{1}{{\dot \phi }}\left(-2\varepsilon H \rho _\phi (1-\frac{\varepsilon }{3})-\frac{\rho _\phi \dot{\varepsilon} }{3}\right),
\end{equation}
this equation by considering Eq.(\ref{28-potential}) can be reexpressed as
 \begin{equation}\label{30-potential --prime}
\frac{{V'(\phi )}}{{V(\phi )}} =  - \frac{1}{{\dot \phi }}\left[\frac{{\dot{\varepsilon}  + 2H\varepsilon (3 - \varepsilon )}}{{3 - \varepsilon }}\right].
\end{equation}
At last by using $H\left( \phi  \right) =  - {\rm{}}\varepsilon {{{H^2}}}/{{\dot \phi }}$ and $\dot V = {\rm{V'}}\dot \phi $ , time evolution of the first slow roll parameter is achieved as
 \begin{equation}\label{31-potential -dot}
\dot \varepsilon  =  - 2H\varepsilon (3 - \varepsilon )\left[1 - {(\frac{{{\varepsilon _{V}}}}{\varepsilon })^{\frac{1}{2}}}\right].
\end{equation}

\subsection{Power spectra of scalar and tensor}
The line element of a spatial FLRW background to bring in account scalar and tensor perturbations is
 \begin{equation}\label{32-perturbed metric}
d{s^2} = (1 + 2A)d{t^2} - 2a(t)({\partial _i}B)dtd{x^i} - {a^2}(t)\Big[(1 - 2\psi ){\delta _{ij}} + 2({\partial _i}{\partial _j}E) + {h_{ij}}\Big]d{x^i}d{x^j},
\end{equation}
where $A$, $B$, $\psi$ and $E$ related to scalar perturbations and $h_{ij}$ is tensor perturbation. If $\psi$ indicates the metric perturbations and $\delta \phi$ related to scalar field perturbation, the curvature perturbation can be defined as
 \begin{equation}\label{33-perturbed curvature}
\mathcal{R} = \psi  + (\frac{H}{{\dot \phi }})\delta \phi.
\end{equation}
By considering linearized Einstein equation one finds
 \begin{equation}\label{34-perturbed curvature equation}
{{\mathcal{R}''}_k} + 2(\frac{{z'}}{z}){{\mathcal{R}'}_k} + c_s^2{k^2}{\mathcal{R}_k} = 0,
\end{equation}
where  $c_s^2$ is square sound speed. Using Mukhanov-Sasaki  variable ${u_k} = z{\mathcal{R}_k}$, the solution of Eq.(\ref{34-perturbed curvature equation}) is
 \begin{equation}\label{35-answer of curvature perturbation}
{{u''}_k} + (c_s^2{k^2} - \frac{{z''}}{z}){u_k} = 0.
\end{equation}
Here $z$ is defined as
 \begin{equation}\label{36-z definition}
z = \frac{{a{{({\rho _\phi } + {P_\phi })}^{\frac{1}{2}}}}}{{{c_s}H}}.
\end{equation}
Based on  \cite{refine}, the expression for scalar and tensor power spectrum can be achieved as
 \begin{equation}\label{37-rho-s}
{\rho _s}(k) = {\Big(\frac{{{H^2}}}{{2\pi \sqrt {{c_s}({\rho _\phi } + {P_\phi })} }}\Big)^2},
\end{equation}
 \begin{equation}\label{38-rho-T}
{\rho _T}(k) = (\frac{8}{{M_{_{Pl}}^2}}){(\frac{H}{{2\pi }})^2}\simeq{\Big(\frac{{2V(\phi )}}{{3{\pi ^2}M_{_{Pl}}^4}}\Big)^{\frac{1}{2}}}.
\end{equation}
In our model ${\rho _s}$ is obtained as
 \begin{equation}\label{39-rho-s01}
{\rho _s}(k) = \frac{1}{{12{\pi ^2}{c_s}}}\frac{{\tilde F(\phi_{0}){V^3}}}{{M_{_{Pl}}^6{{V'}^2}}}.
\end{equation}

\section{Inflation and some typical potentials}
\subsection{Chaotic inflation}
This type of inflation usually is known by considering power law potential
 \begin{equation}\label{40-V-phi}
V(\phi ) = {V_0}{\phi ^n},\,{V_0},\,n > 0.
\end{equation}
For such potentials, we want to obtain the quantity of scalar field for the end of inflation namely ${{\phi _e(N)}}$.  The end of inflation from the slow roll parameters point of view happens when $\varepsilon_V$ grows and identical to unity. Using Eq.(\ref{14-1st slow rool 04}) and above concepts one finds
 \begin{equation}\label{41-phi-e}
\bar{\phi }_{e}\equiv\frac{{{\phi _e}}}{{{M_{Pl}}}} = \frac{n}{{\sqrt {2\tilde F(\phi_{0})} }}.
\end{equation}
It is obvious that, to investigate the accuracy a model we should calculate some important parameters and then compare them with observations. Amongst such parameters one can mention scalar spectral index $n_s$, tensor spectral index $n_T$ and the tensor-to-scalar ratio $\mathfrak{r}$.
From Eqs.(\ref{40-V-phi}) and (\ref{41-phi-e}), $\rho_s$ can be attained as
\begin{equation}\label{42-rho-s}
{\rho _s}(k) = \frac{1}{{12{\pi ^2}{c_s}}}[\frac{{{V_0}\tilde F(\phi_{0})M_{Pl}^{n - 4}}}{{{n^2}}}]{(\frac{\phi }{{{M_{Pl}}}})^{n + 2}}.
\end{equation}
Using definition of $n_s$ and this fact that $\frac{d}{{d\ln k}} =  - \frac{d}{{dN}}$, combination of Eqs.(\ref{39-rho-s01}) and (\ref{41-phi-e}) yields
\begin{equation}\label{43-n-s}
{n_s} - 1 = \frac{{d\ln {\rho _s}(k)}}{{d\ln k}}=\frac{{ - (n + 2)}}{\phi }\frac{{d\phi }}{{dN}}.
\end{equation}
Whereas this equation is based on differentiation of scalar field, $\phi$ should be determined.To obtain scalar field we use definition of e fold number  and Eq.(\ref{20-H Prime}) and it obtained as
\begin{equation}\label{44-phi01}
\frac{\phi (N)}{M_{_{Pl}}} = \frac{{\sqrt {(\frac{n}{2} + 2N)} }}{{\sqrt C }},
\end{equation}
where $C = \frac{{\tilde{F}}}{{n}}$. From this equation and (\ref{41-phi-e}) the initial value of scalar field $\phi_{i}=\phi(N)+\phi_{e}$ is attained as
\begin{equation}\label{45-phi-i}
\bar{\phi}_{i}\equiv\frac{\phi_{i}}{M_{_{Pl}}}=\frac{n}{\sqrt{\tilde F}}(1+\frac{2N}{n}).
\end{equation}
Substituting relation (\ref{44-phi01}) in Eq.(\ref{43-n-s}), $n_s$ is obtained as
\begin{equation}\label{45-n-s-N0}
{n_s} - 1 = \frac{{ - 2(n + 2)}}{{n + 4N}}.
\end{equation}
For more investigation one can use two examples as $n=2$ and $n=4$ in Eq.(\ref{40-V-phi})respectively. For $n=2$ this model can be considered as $m^2 \phi^2/2$ case and for such quantity, $n_s$ is attained as
\begin{equation}\label{46-n-s-n=2}
{n_s} = 1-\frac{{ 4}}{{2N+1}}.
\end{equation}
It should be noted that this result is independent of $\tilde F(\phi_{0}$, hence scalar spectral index for both local and non-local models is identical.
As second case if one consider $n=4$, Eq.(\ref{40-V-phi}) is similar to $\lambda \phi^4/4$ model and $n_s$ for such quantity is as follows
\begin{equation}\label{47-n-s-n=4}
{n_s} - 1 = \frac{{ - 2(n + 2)}}{{n + 4N}}.
\end{equation}
Now we turn our attention to tensor spectral index $n_T$, by substituting power law potential (\ref{40-V-phi}) in Eq.(\ref{38-rho-T}) and using ${n_T} = \frac{{d\ln {\rho _T}(k)}}{{d\ln k}}$, one finds
\begin{equation}\label{48-n-T}
{n_T} = \frac{{ - 2n}}{{4N + n}}.
\end{equation}
As we mentioned above one of the important inflationary parameters is the tensor-to-scalar ratio which by  considering Eqs.(\ref{37-rho-s}) and (\ref{38-rho-T}), it can be attained as
\begin{equation}\label{49-r00}
\mathfrak{r} = \frac{{{\rho _T}(k)}}{{{\rho _s}(k)}} = \frac{{(\frac{2}{{3{\pi ^2}}})(\frac{{{V_0}}}{{M_{Pl}^{n - 4}}}){{(\frac{\phi }{{{M_{Pl}}}})}^n}}}{{\frac{1}{{12{\pi ^2}{c_s}}}\Big[\frac{{{V_0}\tilde F}}{{{n^2}}}M_{Pl}^{n - 4}{{(\frac{\phi }{{{M_{Pl}}}})}^{n + 2}}\Big]}},
\end{equation}
after some manipulation  and using Eq.(\ref{44-phi01}) one finds
\begin{equation}\label{50-tensor-t0-scalar}
\mathfrak{r }= \frac{{16n{c_s}}}{{n + 4N}}.
\end{equation}
We should emphasize our results for ${\rho _s}(k) $ and ${\rho _T}(k) $ are obtained using  two different horizons. For ${\rho _s}(k) $ we consider sound horizon exit and ${\rho _T}(k) $ calculated at horizon exit, albeit due to $H$ is constant during slow roll epoch these two horizons are identical.
The effect of tensor fluctuations on the CMB polarization can be expressed by tensor-to-scalar ratio $\mathfrak{r }$ quantity. Hence one of the main goals of the CMB survey is $\mathfrak{r }$ constraining. Recent data which risen from $WMAP9$ and $STP$ indicate that $\mathfrak{r }<0.13$ and
$\mathfrak{r }<0.11$ at $95\%$ C.L. respectively. Recently based on announcement of $BICEP2$ has detected $B$ modes polarization at the level of $\mathfrak{r }=0.2$. From Eqs.(\ref{48-n-T}) and (\ref{50-tensor-t0-scalar}) one has
\begin{equation}\label{51-tensor-t0-scalar}
\mathfrak{r }= -8n_T c_s,
\end{equation}
which is identical with $\mathfrak{r }$ in non canonical scenario. Therefore the consistency relation for for both non-local and non canonical models differ from the canonical case $\mathfrak{r }= -8n_T. $ As it is mentioned in \cite{refine}, Eq.(\ref{51-tensor-t0-scalar}) emerges as a smoking gun test for the inflationary models examined in \cite{refine}. For three parameters which was obtained for power law potential we compare them with observed quantity which risen from $WMAP9+eCMB+BAO+H_0$ data set  and also Planck data. In addition as it was mentioned the $BICEP2$  data set completed three years data surveying, and their results expressed a constraint on the tensor-to-scalar ratio as $\mathfrak{r }=0.20^{+0.07}_{-0.05}$.
In table {\ref{tab:Moste_P_marginalized}}, the values of spectral indices and tensor-to-scalar ratio for $n=2$ are compared with their observed quantity considering  $WMAP9+eCMB+BAO+H_0$ data set  and also Planck data. In table \ref{tab:Moste_P_marginalized02}, the values of spectral indices and also $\mathfrak{r}$ for $n=3/2$ in comparison to observed quantity are brought. From tables \ref{tab:Moste_P_marginalized} and \ref{tab:Moste_P_marginalized02} it is observed that for $n=2$, $-n_T$ and $\mathfrak{r}$ yield best results in comparison with $n=3/2$ case and also $\bar{\phi}_{i}$ and $\bar{\phi}_{e}$ for $n=3/2$ have acceptable results in comparison to standard models.
\subsection{The exponential potential}
As second case we consider an exponential potential as
\begin{equation}\label{52-V-exponential}
V(\phi ) = {V_0}\exp\Big[- \sqrt {\frac{2}{q}} \frac{\phi }{{{M_{Pl}}}}\Big],
\end{equation}
where $q$ can be appeared in the power of time in scale factor $a(t)=t^q$. By differentiate above equation one obtains
\begin{equation}\label{52-V-prime-exponential}
\frac{dV(\phi )}{V(\phi )} = - \sqrt {\frac{2}{q}} \frac{d\phi }{{{M_{Pl}}}},
\end{equation}
where integration of it from $\phi_{e}$ to $\phi$ gives
\begin{equation}\label{52-phi-exponential}
\frac{\phi_{e}}{\phi} = - \sqrt {\frac{2}{q}} \frac{d\phi }{{{M_{Pl}}}}(\phi_{e}-\phi).
\end{equation}
It is obvious that based on e-folds number definition and Eq.(\ref{52-V-exponential}) one has
\begin{equation}\label{53-N-exponential}
N =  - \frac{\tilde F(\phi_{0})}{{{M_{Pl}}}}\sqrt {\frac{q}{2}} (\phi  - {\phi _e}),
\end{equation}
thus by substituting above relation in (\ref{52-phi-exponential}), one can obtain a definition for $\phi_{e}$ based on $N$, $q$ and $\tilde{F}$ as
\begin{equation}\label{53-phi-bar-exponential}
\bar{\phi}_{e}= \frac{N^{2}}{\tilde F(\phi_{0})N\sqrt{\frac{q}{2}}+\tilde F^{2}(\phi_{0})\sqrt[3]{\frac{q}{2}}}.
\end{equation}
and therefore the initial value for scalar field $\phi_{i}$ could be achieved as
\begin{equation}\label{53-phi-initial-exponential}
\bar{\phi}_{i}= \frac{2N^{2}}{\tilde F(\phi_{0})N\sqrt{\frac{q}{2}}+\tilde F^{2}(\phi_{0})\sqrt[3]{\frac{q}{2}}}-\frac{N}{\tilde F(\phi_{0})}\sqrt{\frac{q}{2}}.
\end{equation}
Also from definition of the first slow roll parameter $n_s$ is obtained as
\begin{equation}\label{55-rho-s-exponential}
{n_s} - 1 =  - \frac{d}{{dN}}\ln {\rho _s}(k)=-\frac{2}{q\tilde F(\phi_{0})}+\frac{N\big(N+q\tilde F(\phi_{0})\big)}{\tilde F^{3}(\phi_{0})\left[ \frac{N}{\tilde F(\phi_{0})}\sqrt{\frac{q}{2}}+\sqrt[3]{\frac{q}{2}}\right]},
\end{equation}
Using Eq.(\ref{38-rho-T}) and (\ref{52-V-exponential}) for the tensor spectral index we find that
\begin{equation}\label{56-rho-s-exponential01}
 - \frac{d}{{dN}}\ln {\rho _s}(k)=-\frac{2}{q\tilde F(\phi_{0})}+\frac{N\big(N+q\tilde F(\phi_{0})\big)}{\tilde F^{3}(\phi_{0})\left[ \frac{N}{\tilde F(\phi_{0})}\sqrt{\frac{q}{2}}+\sqrt[3]{\frac{q}{2}}\right]},
\end{equation}
and from two above equations, the quantities which are obtained for tensor-to-scalar ratio $\mathfrak{r}$ will be brought in {\ref{tab:Moste_P_marginalized03}}.
In table {\ref{tab:Moste_P_marginalized03}} we use the observed quantity for $n_s$, $n_T$ and $\mathfrak{r}$ which risen from Planck data and  $WMAP9+eCMB+BAO+H_0$ data set and compare our results for exponential potential in a non-local mechanism. In addition for different values of $q$ and $\tilde F$, both $\bar{\phi}_{i}$ and $\bar{\phi}_{e}$ are obtained.
\newpage
\renewcommand{\arraystretch}{1.1}
\begin{table}
\begin{center}
\rowcolors{1}{Lyellow}{white}
\begin{tabular}{|c|c|c|c|c|}
\hline\hline
$N$ & $60$ &  $55$ & $65$ &  $Observed$ \\
\hline
$n_{s}$ & $0.9670$ &  $0.9639$ & $0.9694$ & $<0.9675$ \\
\hline
$-n_{T}$ & $0.016$ &  $0.018$ & $0.015$ &  $<0.016$  \\
\hline
$\mathfrak{r}$ & $0.13$ &  $0.14$ & $0.12$ &  $<0.13$  \\
\hline
$\bar{\phi}_{i}$ & $4.93$ &  $4.73$ & $5.13$ &  $-$ \\
\hline
$\bar{\phi}_{e}$ & $0.44$ &  $0.44$ & $0.44$ &  $-$ \\
\hline\hline
\end{tabular}
\end{center}
\caption{\small{For $n=2$ in chaotic inflation, scalar and tensor spectral indices are considered for three different quantities of the $N$. The observed quantity of $n_s$ is brought from Planck data, $-n_T$ is risen from $WMAP9+eCMB+BAO+H_0$  data set and $\mathfrak{r}$ is considered from $WMAP9$. It should be stresses for this example we consider $\tilde F=10.$
}}\label{tab:Moste_P_marginalized}
\end{table}

\renewcommand{\arraystretch}{1.1}
\begin{table}
\begin{center}
\rowcolors{1}{Lyellow}{white}
\begin{tabular}{|c|c|c|c|c|}
\hline\hline
$N$ & $60$ &  $55$ & $65$ &  $Observed$ \\
\hline
$n_{s}$ & $0.9712$ &  $0.9684$ & $0.9732$ & $<0.9675$ \\
\hline
$-n_{T}$ & $0.012$ &  $0.013$ & $0.011$ &  $<0.016$  \\
\hline
$\mathfrak{r}$ & $0.099$ &  $0.1$ & $0.091$ &  $<0.13$  \\
\hline
$\bar{\phi}_{i}$ & $4.30$ &  $4.80$ & $4.43$ &  $-$ \\
\hline
$\bar{\phi}_{e}$ & $0.33$ &  $0.33$ & $0.33$ &  $-$ \\
\hline\hline
\end{tabular}
\end{center}
\caption{\small{For $n=3/2$ in chaotic inflation, scalar and tensor spectral indices are considered for three different quantities of the $N$. The observed quantity of $n_s$ is brought from Planck data, $-n_T$ is risen from $WMAP9+eCMB+BAO+H_0$  data set and $\mathfrak{r}$ is considered from $WMAP9$. It should be stresses for this example we consider $\tilde F=10.$
}}\label{tab:Moste_P_marginalized02}
\end{table}
\renewcommand{\arraystretch}{1.1}
\begin{table}
\begin{center}
\rowcolors{1}{Lyellow}{white}
\begin{tabular}{|c|c|c|c|c|c|}
\hline\hline
$N$ & $60$ &  $55$ & $65$ &  $Observed$ & $q=2, \tilde F=32$\\
\hline
$n_{s}$ & $0.9962$ &  $0.9957$ & $0.9965$ & $<0.9675$ & $"$ \\
\hline
$-n_{T}$ & $0.037$ &  $0.004$ & $0.003$ &  $<0.016$ & $"$ \\
\hline
$\mathfrak{r}$ & $0.25$ &  $0.25$ & $0.25$ &  $<0.13$ & $"$ \\
\hline\hline
\end{tabular}
\end{center}
\caption{\small{For exponential potential, scalar and tensor spectral indices are considered for three different quantities of the $N$. The observed quantity of $n_s$ is brought from Planck data, $-n_T$ is risen from $WMAP9+eCMB+BAO+H_0$  data set and $\mathfrak{r}$ is considered from $WMAP9$.
}}\label{tab:Moste_P_marginalized03}
\end{table}
\renewcommand{\arraystretch}{1.1}
\begin{table}
\begin{center}
\rowcolors{1}{Lyellow}{white}
\begin{tabular}{|c|c|c|c|c|c|}
\hline\hline
$N$ & $60$ &  $55$ & $65$ &  $Observed$ & $q=2, \tilde F=60$\\
\hline
$n_{s}$ & $0.9958$ &  $0.9951$ & $0.9965$ & $<0.9675$ & $"$ \\
\hline
$-n_{T}$ & $0.004$ &  $0.0043$ & $0.0034$ &  $<0.016$ & $"$ \\
\hline
$\mathfrak{r}$ & $0.13$ &  $0.13$ & $0.13$ &  $<0.13$ & $"$ \\
\hline\hline
\end{tabular}
\end{center}
\caption{\small{For exponential potential, scalar and tensor spectral indices are considered for three different quantities of the $N$. The observed quantity of $n_s$ is brought from Planck data, $-n_T$ is risen from $WMAP9+eCMB+BAO+H_0$  data set and $\mathfrak{r}$ is considered from $WMAP9$.
}}\label{tab:Moste_P_marginalized04}
\end{table}
\renewcommand{\arraystretch}{1.1}
\begin{table}
\begin{center}
\rowcolors{1}{Lyellow}{white}
\begin{tabular}{|c|c|c|c|c|c|}
\hline\hline
$N$ & $60$ &  $55$ & $65$ &  $Observed$ & $q=1, \tilde F=30$\\
\hline
$n_{s}$ & $0.9958$ &  $0.9962$ & $0.9954$ & $<0.9675$ & $"$ \\
\hline
$-n_{T}$ & $0.0041$ &  $0.0037$ & $0.0045$ &  $<0.016$ & $"$ \\
\hline
$\mathfrak{r}$ & $0.16$ &  $0.16$ & $0.16$ &  $<0.13$ & $"$ \\
\hline\hline
\end{tabular}
\end{center}
\caption{\small{For exponential potential, scalar and tensor spectral indices are considered for three different quantities of the $N$. The observed quantity of $n_s$ is brought from Planck data, $-n_T$ is risen from $WMAP9+eCMB+BAO+H_0$  data set and $\mathfrak{r}$ is considered from $WMAP9$.
}}\label{tab:Moste_P_marginalized05}
\end{table}
\section{Conclusion}
The inflationary universe within the framework of  considering a non-local scalar field have discussed. Two well known models of inflation includes of power law and exponential potentials to investigate the effects of non-locality on the inflationary parameters have been considered . Also some important parameters includes of slow roll parameters, scalar and tensor power spectra, spectral indices, the tensor-to-scalar ratio and so on for chaotic and exponential inflationary scenarios have been calculated. To investigate inflationary attractors  Hamilton-Jacobi formalism as a suitable  way to study the effect of perturbation based on e-folding number $N$ have been investigated. Also free parameters of the model with observed values by considering Planck data, $WMAP9+eCMB+BAO+H_0$ data set in addition to $BICEP2$ data surveying have been compared. It has shown that our theoretical results are in acceptable agreement in comparison to observed values. Also for different potentials which were studied in this work, we can emphasise the tensor-to-scalar ratio for exponential potential, by considering $BICEP2$ is in better agreement in comparison with chaotic inflation. Also the results of comparison between theoretical results and observations can be summarized as bellow:\\
In table {\ref{tab:Moste_P_marginalized}}, the values of spectral indices and tensor-to-scalar ratio for $n=2$ have compared with their observed values using  $WMAP9+eCMB+BAO+H_0$ data set  and also Planck 2013 data. In table \ref{tab:Moste_P_marginalized02}, the values of spectral indices and also $\mathfrak{r}$ for $n=3/2$ in comparison to observed quantity have brought. From tables \ref{tab:Moste_P_marginalized} and \ref{tab:Moste_P_marginalized02} it is observed that for $n=3/2$, $-n_T$ and $\mathfrak{r}$ yield best results in comparison with $n=2$ case. In addition the quantities of scalar field which related to the begin and end of inflation have been calculated, and it was find out the $n=3/2$ case has better results in comparison to $n=2$ model. In tables {\ref{tab:Moste_P_marginalized03}}, {\ref{tab:Moste_P_marginalized04}} and {\ref{tab:Moste_P_marginalized05}} the observed quantity for $n_s$, $n_T$ and $\mathfrak{r}$ which risen from Planck data and  $WMAP9+eCMB+BAO+H_0$ data sets have considered and  our results for exponential potential in a non-local mechanism  have compared with them.
It was observed that $\mathfrak{r}$ is in best agreement for $q=2$  and $\tilde F=60$, also the quantities which were obtained for $n_{s}$ had some differences with observations. It was interesting to note that, if one consider the result of $BICEP2$ for $\mathfrak{r}$ ,$0.2,$ $q=2$  and $\tilde F=32$, produced better result in comparison to other quantities for $q$ and $\tilde F$.
\section{ Acknowledgment }
The  authors  thank A.  Aghamohammadi and S. W. Rabiei  for  very  useful  and  stimulating  discussions.


\end{document}